\documentclass[12pt]{article}

\def \g {\gamma}
\def \L {\Lambda}

\newcommand{\rf}[1]{(\ref{#1})}
\def \ci {\cite}
\def \del{\partial}

\def \foot{\footnote}
\def \p {\phi}
\def \la {\label}
\def \bi {\bibitem}
\def \G {\Gamma}
\def \const{{\rm const}}
\def \vt {\vartheta}
\def \S {{\cal K}}

\def \ads {$AdS_5 \times S^5$ } 
\def \ov {\over}
\def \tx {\tilde x}
\def \td {\tilde}

\def \k {\kappa}

\def \ha {{1 \ov 2}}
\def  \diag {{\rm diag}}

\def\np {{  Nucl. Phys. }}
\def \pl {{  Phys. Lett. }}

\def \prl {{  Phys. Rev. Lett. }}

\def \ep {\epsilon}

\newcommand{\be}{\begin{equation}}
\newcommand{\ee}{\end{equation}}
\newcommand{\bea}{\begin{eqnarray}}
\newcommand{\eea}{\end{eqnarray}}
\newcommand{\eps}{\epsilon}

\newcommand{\Pp}{{\cal P}_+}
\newcommand{\Pm}{{\cal P}_-}
\newcommand{\Ppm}{{\cal P}_\pm}

\newcommand{\s}{\sigma}

\newcommand{\ksym}{$\kappa$-symmetry }

\begin{document}
\begin{flushright}
\hfill{SU-ITP-98/49}\\
\hfill{Imperial/TP/97-98/65   }\\
\hfill{hep-th/9808088}\\
\hfill{August 1998}\\
\end{flushright}

\vspace{20pt}

\begin{center}
{\large {\bf
Simplifying  Superstring  Action   on   $AdS_5\times S^5$   
}}

\vspace{40pt}

{\bf Renata Kallosh$^{1,a}$ and Arkady A. Tseytlin$^{2,b}$}

\vspace{10pt}
{\it $^1$ Department of Physics

Stanford University

Stanford, CA 94305-4060, USA}

\vspace{10pt}
{\it $^2$  Blackett Laboratory

Imperial College

London, SW7 2BZ, U.K.}

\vspace{60pt}

\underline{ABSTRACT}

\end{center}

 Type IIB string action on $AdS_5\times S^5$ 
constructed in hep-th/9805028 is  put into a form  where it becomes 
quadratic  in fermions. This is achieved by  
performing 2-d duality
(T-duality)   on  the  action in which kappa-symmetry was fixed in the  
Killing gauge
in hep-th/9808038.
We discuss some properties and possible applications of the resulting action.

{\vfill\leftline{}\vfill
\vskip  30pt
\footnoterule
\noindent
{\footnotesize
$\phantom{a}^a$ e-mail: kallosh@physics.stanford.edu. }  \vskip  -5pt

\noindent
{\footnotesize
$\phantom{b}^b$ e-mail: a.tseytlin@ic.ac.uk. Also at Lebedev Physics Institute,
Moscow. }  \vskip  -5pt


\pagebreak
\setcounter{page}{1}

1. \  The  classical  superstring action of Green-Schwarz type \cite{GS}  was 
recently constructed in the  non-trivial 
maximally supersymmetric  $D=10$ type IIB  supergravity  vacuum  \ci{SH}
 (which is also  the near horizon  space of the D3 brane \ci{GT}), 
i.e. in the 
$AdS_5 \times S^5$  background  \cite{Tseytlin,KRR,TTM}.
 This  
action has
local \ksym   and 2-d reparametrization symmetry.  By gauge-fixing   
\ksym one can 
reduce the number of fermions  by 1/2  to match the number of physical  
bosonic
and fermionic degrees of freedom.   The gauge-fixing  of \ksym  was  
performed
in \cite{KR} developing  the proposal \cite{SuperKilling} and
 the  
action was
found which has  terms at most
 quartic in fermions.\footnote{A similar action 
was found  in 
\cite{Pesando} using  supersolvable
 ({\it Ssolv})  algebra approach \ci{SS}.  That  action
will be qualified as  a  gauge-fixed action if the relation between the
gauge-fixing and {\it Ssolv} algebra approach is clarified. 
It is   
not quite
clear whether the actions in \cite{KR} and in \cite{Pesando}  
actually agree given also that
 different choices of bosonic (and fermionic)
 coordinates
were used:  Cartesian 
and horospherical
 in \cite{KR} and projective coordinates on $S^5$  in
\cite{Pesando}.
}
The  special $\k$-symmetry gauge 
 using the  projector parallel to D3-brane directions
  allowed to substantially  reduce the  power of fermionic terms in the  
action.
Still, in contrast to the Green-Schwarz  action in the light-cone gauge
(in flat space or on a group manifold), 
the  resulting 
action was not
quadratic in fermions. 

One of the  purposes of this  paper is to describe  
a simple  transformation that  relates  
the gauge-fixed action  to an action  which is   
{\it quadratic}  in fermions. 
This transformation is the  2-d scalar-scalar  duality applied to 
the  string coordinates parallel to the  `D3-brane' directions.
The corresponding target space transformation is T-duality 
along the 4 world-volume D3-brane directions. It 
 transforms the near-core D3-brane background (i.e. $AdS_5 \times S^5$)
into the near-core D-instanton background \ci{GGP}
smeared in the 4 directions.
This T-dual background  has  conformally flat
$D=10$ string-frame metric which, remarkably, is again equivalent to 
the   $AdS_5 \times S^5$ metric  (as can be seen 
by  the coordinate transformation 
$y \to 1/y$ in the radial direction).\foot{Note that the 
near-core string-frame 
metric
of the  {\it localized} (not smeared)   D-instanton background  is flat \ci{TT,BB}.}
 In addition, there 
are non-vanishing  
 dilaton   and  Ramond-Ramond  scalar   backgrounds.\foot{The   
 $AdS_5 \times S^5$ metric  can  be completed to a solution of type IIB supergravity  by  either  the  RR 4-form background
or by the  T-dual dilaton  and  RR 0-form background.}

  The resulting   dual 
action (which,  as usual, 
 is  expected  to represent an equivalent 2-d conformal theory, at least 
in the case of toroidal compactification of D3-brane directions)
can thus  be interpreted as an action of a fundamental string 
propagating  in  the near-core region of the smeared D-instanton
background. 
 The drastic simplification of the fermionic part of the action 
may be  related to the fact  that the  D-instanton background
 has flat $D=10$ Einstein-frame metric.

The fermionic part of the action  takes the form 
$ \bar \theta A \theta , $  where 
$A$ is a  first-order differential operator,  $A\sim \del X \del $.
Remarkably, $A$ does not depend on the 2-d metric as 
the fermionic term is of WZ type. Since  the
\ksym gauge is   already  fixed, $A$   should be
non-degenerate  for generic  string  background.   
We shall show that 
invertibility of the operator $A$ puts certain 
 constraints on the  
properties of the bosonic string coordinate $X$  
background.   

Having quadratic  fermionic action  should be 
 important  for solution of the corresponding  2-d conformal model.
 At the classical level, one  may  be able to 
 solve the fermionic equations of motion explicitly.
At the quantum level, one can   integrate out  the 
fermions,    obtaining the 
effective action
 $S_{eff}(X)  = S (X) - \ha\ $ln det $A(X)$\  for the bosonic coordinates
 $X$.
We expect that the  fermionic determinant  will
 induce a WZ-type term
(cf. \ci{WI}),    explaining  how the presence 
of the fermionic couplings representing  a  non-trivial RR background 
makes the symmetric space \ads  sigma model 
 conformal at the quantum level. That  would   
  provide a
non-perturbative 
demonstration  of the conformal invariance of 
the \ads superstring  model, 
complementing  the all-order perturbative 
conformal invariance  proof  given  in \ci{Tseytlin,TTT}. 
While the  bosonic \ads sigma-model is   classically 
integrable
but not solvable and
 not conformal  at the quantum level, 
one may   expect  that  the 
theory defined by the  superstring action or
 $S_{eff}(X)$  will  have properties  which are  similar  to those 
of the group space sigma model  
 (WZW theory), and, in particular,  may 
be explicitly solvable as a 2-d theory.

One possible application of the  superstring 
action that we shall briefly 
discuss below is to the computation  of the 1-loop correction 
to the semiclassical value  of the Wilson factor  
in \ci{MR}.  It is plausible  that the L\"uscher $1/L$  term 
in the effective potential  is non-vanishing in this  
case 
(cf. 
\ci{GO}).


2.\   We start with the classical \ads  action obtained in the
 closed form via a 
 supercoset construction \cite{Tseytlin,KRR,TTM} ($2\pi \alpha'=1$) 
\begin{eqnarray}
S =-\frac{1}{2}\int d^2\sigma\ \left(\sqrt{-g} \, g^{ij}
 L_i^{\hat a} L_j^{\hat a} +  4 i \eps^{ij}\int_{0}^1 ds\ 
 L_{is}^{\hat a} \S^{IJ} \bar \Theta^I \Gamma^{\hat a} L_{is}^{J}
 \right)
\ .
\label{action}
\end{eqnarray}
Here  $\S^{IJ}\equiv$ diag$(1,-1)$,  \ $I,J=1,2$  and  $\hat
a=(a,a')=(0,...,4,5,...,9)$. The invariant 1-forms
$L^I=L^I_{s=1}, \ L^{\hat a}=L^{\hat a}_{s=1}$ 
are given by
\begin{equation}
L_s^{ I} = \bigg({\sinh \left({s\cal M}\right) \over {\cal  
M}}
D\Theta
\bigg)^{I}\ , \  \quad L_s^{\hat a }=e^{\hat a }_{\hat m} (X) dX^{\hat m}  
- 4 i \bar
\Theta^I\Gamma^{\hat a}
\bigg({
\sinh^2  \left({\ha s\cal M}\right) \over {\cal M}^2} D\Theta \bigg)^I\ , 
\label{LI}
\end{equation}
where  $X^{\hat m}$ and $ \Theta^I$ are the bosonic and fermionic
superstring coordinates and 
\be
({\cal M}^2)^{ IL}=  \epsilon^ {IJ} (-\gamma^{ a} \Theta^{J} \bar
\Theta^L \gamma^{ a} + \gamma^{ a'}
\Theta^{J} \bar \Theta^L \gamma^{ a'} ) + {1\over 2}
\epsilon^{KL} (\gamma^{ab} \Theta^I \bar \Theta^K \gamma^{ab}
-\gamma^{a'b'} \Theta^I \bar \Theta^K \gamma^{a'b'})
\ , \label{msquare}
\ee
\begin{equation}
(D\Theta)^I = \bigg[  d +{1\over 4}(\omega^{ab} \gamma_{
ab}  + 
\omega^{a'b'}  \gamma_{
a'b'}  ) \bigg]
  \Theta^I  -{1\over 2} i\epsilon^{IJ}(  e^{ a} \g_a  + i e^{a'} \g_{a'} 
)\Theta ^J \ . \label{DTheta}
\end{equation}
The Dirac matrices are split in the  `5+5'  way,  
$\Gamma^a= \gamma^a \times 1 \times  \s_1, \ \
\Gamma^{a'}= 1 \times \gamma^{a'} \times \s_2,$
where $\s_k$ are Pauli matrices
(see \ci{Tseytlin,TTM} for details on notation). 

Let us review  the  \ksym 
 gauge fixing of this action performed in \cite{KR}.
 We shall use the `D3-brane adapted'  or `4+6'
bosonic coordinates $X^{\hat m}
=(x^p,y^t)$ 
in which the \ads metric  is  split into  the parts 
parallel and transverse to the D3-brane directions
(we take the radius parameter   to be $R=1$)  
\be
ds^2 = y^2 dx^p dx^p  + {1 \ov y^2} dy^t dy^t \ ,\ \ \ \  
\ \ \ \ \   y^2 \equiv y^t y^t \ , 
\la{met}
\ee
where 
$ p=0,...,3,\   t= 4,...,9 $. In what follows the 
contractions of the indices $p$ 
is understood  with Minkowski metric and indices $t$  -- with Euclidean metric. 
The \ksym gauge is fixed   using the `parallel to  D3-brane'
  $\Gamma$-matrix
projector\foot{This  projector  is hermitean and  anticommutes with $\Gamma^0$
so that $ \bar \Theta_+ =  
\Theta^\dagger \Pp
\Gamma^0 = \bar \Theta \Pm$, \  
$
\bar \Theta \equiv  \Theta^\dagger \Gamma^0$. 
For two spinors $\Theta$ and $\Psi$ one has 
$
\bar \Theta_{+}^I  \Gamma^{\hat a} \Psi_+^I =
 \Theta^\dagger  \Gamma^0 \Pm
 \Gamma^{\hat a} \Pp \Psi
$. Since $\Gamma_{0123}$ anticommutes with  
$\Gamma^p$,  this
expression does not vanish
 only in  the `parallel' directions, i.e.
$\bar \Theta_{+}  \Gamma^{p} \Psi_+ \neq 0, \  \bar \Theta_{+}   
\Gamma^{t}
\Psi_+=0.$ 
For example, only the `parallel' part (${\hat a}=p$)  of 
$\bar \Theta^I_+ \G^{\hat a} \del \Theta^I_+$ 
remains in this gauge in the `kinetic' part of the
flat-space superstring action.
At the same time, only the `transverse' part (${\hat a}=t$)
survives in the combination
$ \S^{IJ} \bar \Theta_{+}^I  \Gamma^{\hat a} \del  \Theta_+^J $
which appears in the WZ term in the string action.}
\be 
\Theta_-^I=0 \ , \ \ \ \ \ 
\Theta_{\pm}^I\equiv  \Ppm^{IJ} \Theta^J \ , \qquad \Ppm^{IJ}=\frac{1}{2}\left(\delta^{IJ} \pm
\Gamma_{0123}\eps^{IJ}\right)\  , \ \ \ \ \  \Pp \Pm=0
\ . 
\la{gau}
\ee
In `5+5' coordinates ($x^a=(x^p,x^4=y)$ and $\xi^{a'}$ coordinates on $S^5$)
one finds   that 
($\G_{0123} = i \g_4 \times 1 \times 1$, \ $\omega^{p4}=e^p$)
\begin{equation}
(D\Theta)^I  = 
 \bigg[ \delta^{IJ} (  d +{1\over 4} \omega^{a'b'} \gamma_{a'b'}) 
  + {1\over 2} \ep^{IJ} (  e^{ a'} \g_{a'}  - i e^4 \g_4 ) 
+ {1\over 2} e^p  \g_p \g_4 \Pm^{IJ} \bigg] \Theta^J \ . 
\label{DTHETA}\end{equation}
Using that  
the $S^5$ part  of the covariant derivative 
satisfies  $  D_5^{IJ} \equiv   \delta^{IJ} 
 (  d +{1\over 4} \omega^{a'b'} \gamma_{a'b'})  + {1\over 2} \ep^{IJ} 
 e^{ a'} \g_{a'} = (\L d \L^{-1})^{IJ}, \  (D_5)^2=0$, 
where  the spinor matrix $\L^{IJ}=\L^{IJ}(\xi)$ is a function of the $S^5$ coordinates,\foot{The same  matrix 
appears in the expression for the  Killing spinors on $S^5$ 
 \ci{pop}.}
 one finds  that 
$(D\Theta_+)^I$ can be 
   written as 
\begin{equation}
D\Theta_+ = (d - \ha d \log y + \L d \L^{-1}) \Theta^I_+  =
 y^{1/2} \L \ d \  ( y^{-1/2}\L^{-1} \Theta_+)    \ . \la{red}
\end{equation}  
Eq. \rf{red}  suggests 
to make  the change of the fermionic variable
$\Theta \to \theta$ 
\be
\Theta^I_+  = y^{1/2}\L^{IJ}(\xi) \ \theta^J_+ \ , \ \ \ \ \ \
\Pm^{IJ}  \theta_+^J =0 \ ,  \ \ \ \ \ \ 
D\Theta_+^I = y^{1/2} \L d \theta_+^I \ .  \la{cha}
\ee
If we  further   transform  from the coordinates  $(y,\xi^{a'})$
to the 6-d Cartesian coordinates $y^t$ in \rf{met}, $y^t= {y\ov \sqrt{1 + \xi^2}} (1, \xi^{a'})$, 
that would effectively absorb the matrix $\L$
into an $SO(6)$  spinor rotation.\foot{This may be interpreted 
as a 
 rotation    from $(y,0,0,0,0,0)$
to a generic  6-vector $y^t$  by the $SO(6)$ transformation 
parametrised by $S^5$ angles or  by the 
 unit vector $\hat y^t= { y^t \ov y}$.
In the Cartesian coordinates $y^t= y \hat y^t $ 
the  6-d  part of the covariant derivative has the form
$D_6^{IJ}= 
\delta^{IJ} 
 (  d +{1\over 2}  \gamma_{st} \hat y^s d\hat y^t)   
+ {1\over 2} \ep^{IJ} 
  \g_{t}  (d\hat y^t  + \hat y^t d\log y)$.
}
This simplification  is suggested
\ci{SuperKilling}
by   the form of the
  Killing spinors  in   \ads space viewed as the 
near-horizon D3-brane 
background.  In particular, writing the 10-d  covariant 
derivative  (including the  Lorentz connection 
and 5-form terms)
in the  `4+6' coordinates in \rf{met}
 one learns \ci{KKU} that  when  
acting on  the constrained spinor
$\Theta_+$ 
it becomes simply 
$D\Theta_+^I = y^{1/2}  d \theta_+^I, \ \  
\theta_+^I \equiv y^{-1/2} \Theta_+^I$.

As a result, ${\cal M}^2 D\Theta_+=0$  \ci{KR} 
and the fermionic sector of the  action  reduces  only to terms 
 quadratic   and quartic  in $\theta_+^I$.  Using 
$\Pm^{IJ}  \theta_+^J =0$ to 
eliminate $\theta^2_+$ in  favour of
 \be \theta^1_+ \equiv \vt   \  \ee
one finds that the \ksym gauge-fixed 
  string action in $AdS_5
\times S^5$ background \rf{action}   
expressed in terms of  the bosonic coordinates  $X^{\hat m} 
=(x^p, y^t)$ and 
the 
{\it single}   $D=10$ Majorana-Weyl spinor  $\vt$
takes the following simple 
 form\foot{Let us summarize the  main steps in the 
derivation of this action. 
One way to  argue  that the string action simplifies 
 in the `4+6'   cartesian coordinates is  to 
 start with the `5+5' or `direct product'  formulation  \ci{Tseytlin}
where one has the  underlying superalgebra  and supercoset space
construction 
 and  thus  can find the explicit 
solution  \ci{KRR} 
for  the supervielbein in terms of the ${\cal  M}^2$ matrix \rf{msquare}\ 
(the existence of such solution is not manifest in the cartesian coordinates where the 10-d metric  \rf{met} 
does  not look  like a direct product).  
One is then able to argue \ci{SuperKilling,KR}
that  after fixing  the `D3-brane' \ksym  gauge
 the  supergeometry   simplifies so that  the  vielbein   contains
$\theta^2$ terms at most.  Hence  the string action should
involve only terms of $n=0,2,4$ power in fermions. 
The structure  of the `kinetic' term in the action is obvious, while 
to  fix  the form of the WZ term 
it is useful to 
 switch  from  the `5+5' to the  manifest 10-d spinor notation.
The  3-form  that appears in the WZ term is \ci{Tseytlin}
$
 {\cal H}= L^{\hat a} \wedge  (\bar L^1 \wedge   \Gamma^{\hat a} 
 d L^1  - \bar L^2 \wedge    \Gamma^{\hat a} d L^2) $
and since in the D3-brane gauge  for 
 the cartesian coordinate metric \rf{met} $
    L^I = y^{1/2} d (y^{-1/2} \Theta^I )$  \ci{KKU,KR}
we learn 
that  the bracket in ${\cal H}$  is   essentially  the same 
as  in  the flat space 
 case and 
  is non-vanishing only for $\hat a=t$, i.e. for the transverse  6-space indices.
Using that $L^t =y^{-1} dy^t$  one  finishes with  the above  simple 
`flat-space'  form of the  WZ term.}
\begin{eqnarray}
S =-\frac{1}{2}\int d^2\sigma\ \biggl[\sqrt{-g} \, g^{ij}&& \hspace{-0.7cm}
\left(y^2(\partial_i x^p
- 2 i \bar
\vt \Gamma^{p} \partial_i\vt)(\partial_j x^p - 2 i \bar
\vt \Gamma^{p} \partial_j \vt) +\frac{1}{y^2} \partial_i y^t
\partial_j y^t \right) \nonumber \\ &&
 +\  4 i \eps^{ij} \partial_i y^t \bar \vt \Gamma^t
\partial_j\vt \ \biggr] \ . 
\label{SA}
\end{eqnarray}
The $\Theta \Theta  \del X\del X$ terms representing 
the coupling to the  RR background 
   present \ci{Tseytlin} in  the original action \rf{action}
are now `hidden'  in the $\bar \vt \del \vt \del  X $ terms
because of    the redefinition  made in \rf{cha}.


The same action but without $y^2$ and $1/y^2$ factors
is found by fixing the  \ksym   gauge \rf{gau}
in the flat-space type IIB  Green-Schwarz action. 
This `D3-brane'  gauge    breaks the  $SO(1,9)$ Lorentz invariance
of the action 
to $SO(1,3) \times SO(6)$, i.e.  distinguishes between the 
4 `parallel' and 6 `transverse' coordinates.
In particular, only the latter ones ($y^t$) 
survive  in the WZ term. This special 
property of the  gauge  \rf{gau} 
turns out to be crucial for the   observation below  that 
the fermionic terms in the action can be  put in a  much 
simpler 
 {\it quadratic}  
  form 
by  making 2-d duality transformation  of the `parallel'
coordinates $x^p$.


3. \  
 Let us  now perform 
the  2-d duality transformation  of the four  $x^p$  coordinates. 
As usual, this is done by putting the action in the first-order 
form by  introducing  the `momenta' (Lagrange multipliers) $P^p_i$ 
\begin{eqnarray}
S =- { 1 \ov 2} 
\int d^2\sigma\ \biggl[\sqrt{-g} \, g^{ij}&& \hspace{-0.7cm}
\left(- {1\over y^2} P^p_i P^p_{j}  + 2  P^p_i (\partial_j x^p - 2 i \bar
\vt \Gamma^{p} \partial_j \vt) +  \frac{1}{ y^2} \partial_i y^t
\partial_j y^t \right) \nonumber \\ &&
+ \ 4 i \eps^{ij} \partial_i y^t  \bar \vt \Gamma^t
\partial_j\vt \biggr] \ . 
\label{SimpleAction1}
\end{eqnarray}
Integrating out $x^p$  and solving the resulting constraint on $P^p_i$
  as 
\be
\sqrt{-g} g^{ij} P_j^p = \epsilon^{ij}\del_j \tx^p  \ , \ee
we finish with the dual action 
$$
\td S = - { 1 \ov 2} \int d^2\sigma\ \bigg[\sqrt{-g} g^{ij} \ 
\frac{1}{y^2}(\partial_i \tx^p \partial_j \tx^p +  \partial_i y^t
\partial_j y^t)  $$
\be 
+\  4i \eps^{ij} \bar
\vt ( \del_i \tx ^p \Gamma^{p} + 
\partial_i y^t  \Gamma^t ) 
 \partial_j \vt \bigg]  \ . 
\la{rsa}
\ee
At the quantum level,  the 2-d duality is accompanied \ci{BU} by the dilaton 
term  \ci{FT}  which should be added to the dual action \rf{rsa}
to preserve its conformal invariance,
\be 
\Delta \td S = { 1 \ov 4\pi } \int d^2 \sigma \sqrt{-g} R^{(2)} \phi (X) \ , 
\ \ \ \ \ \  \phi = \phi_0  - 4 \ln y  \ . 
\la{dil}
\ee  
A   remarkable property of  the action 
\rf{rsa} is not only that  its  fermionic 
part  is quadratic in $\theta$ but also that
 it  
does not depend on 2-d metric, i.e. is given by a WZ type term.
 This 
WZ term is  linear  in the  bosonic coordinates, i.e. 
has formally  the same  form as in flat  target space. 
A  somewhat surprising conclusion is that  adding this fermionic 
term to the bosonic  symmetric space \ads  sigma model 
action  (with dilaton term  \rf{dil} also  included) should 
 give  a conformally invariant 2-d theory! \foot{Let us mention also that  written in terms 
of the variable $\rho =  \ha \ln y$ the bosonic part of the 
action \rf{rsa},\rf{dil} is very similar to the one discussed 
in \ci{AP}: the  background  metric and dilaton are
$ds^2 = d\rho^2 + e^{-2\rho} d\tx^p d\tx^p  + d\Omega_5^2, \ \
\phi = \p_0 - 8 \rho$.}

Let us note that  if we  would start with   the `5+5' 
form of the action  in which the WZ term has  a more complicated 
structure  depending 
on the  $S^5$ spinor matrix $\L(\xi)$ in \rf{cha}
(which drops out of the `kinetic' term in \rf{SA}), 
 that would not affect the 2-d duality transformation 
step, and the resulting  dual action will still be quadratic in fermions.

The duality has  partially
 restored  a  `symmetry' between 
 `parallel' and `transverse' coordinates. 
Starting with  the flat-space analogue  of \rf{SA}
and performing the same duality transformation 
one would get \rf{rsa} without the $1/y^2$ factor, i.e.
obtain  indeed 
 the 
$SO(1,9)$ invariant action for $\td X^{\hat a} = (\td x^p, y^t)$
and $\vt$
\be
\td S_{flat}  = - { 1 \ov 2} \int d^2\sigma\ \bigg(\sqrt{-g} g^{ij} \partial_i \td X^{\hat a}  \partial_j \td X^{\hat a} \ + \ 
 4i \eps^{ij}    \del_i \td X^{\hat a}  \
 \bar \vt   \Gamma^{\hat a }  
 \partial_j \vt \bigg) \ . 
\la{rsaf}
\ee
This looks  like the type I (or heterotic)  flat-space 
string action  but without 
the  standard fermionic terms complementing 
$\del X$  in the `kinetic'  part of 
the action (as a result, the  $\kappa$-symmetry  is broken, as, of course,
 should be in the present type IIB  gauge-fixed  theory).\foot{Note that  fixing the light-cone \ksym   gauge 
$\G^+ \theta=0$  in the flat-space IIB action
leads to  the  
action ($\int \del X^+ \bar \vt   \Gamma^{- }  
 \partial \vt$)
  which is similar to  \rf{rsaf}
but is not 10-d Lorentz-invariant and has  the  
fermionic term   which depends on the 2-d metric, 
while in  \rf{rsaf} the fermionic term 
  has purely `topological'  WZ structure.}
We  would arrive at exactly the same flat-space dual 
 action \rf{rsaf}
had we  started  with the flat-space IIB action, used 
 the  `Dq-brane'  combination
$\G_{0...q}$  ($q=$odd) instead of 
$\G_{0123}$ in \rf{gau}
 and   dualized the `parallel'
coordinates $x^p, \  p=0,...,q$.  However, that 
procedure
 would  no longer generalize to curved \ads 
space unless $q=3$:   the  form of the \ads background \rf{met} 
prefers  the `D3-brane' gauge  choice \rf{gau}.

As was already mentioned above,    the fact 
that 2-d duality simplified  the structure of the 
fermionic terms is related 
to the  key property of the gauge-fixed action \rf{SA}
or its flat-space  counterpart:   only part  
of the bosonic 
 coordinates  (`transverse' ones) appear in the WZ term.  For example, 
the standard type I  superstring  action  which has  similar   form
of a sum of a `kinetic' (2-d metric dependent)
 and a  WZ term,  i.e. 
$ \int ( \del X - \bar \theta \del \theta)^2 + 
i d X \wedge \bar \theta d \theta$, \ 
{\it preserves} its form under 2-d duality  
applied to   any of the  coordinates  $X^{\hat a}$, 
 as all of them 
  enter both  the  first
 and the second term in that action.

The  dual action  \rf{rsa}
can be interpreted
as describing  the fundamental string propagating  in  the background 
representing 
the  near-core region of the  D-instanton 
smeared in the 4 directions $\tx^p$. This 
background  
 is T-dual\footnote{Note that T-duality along the isometric 
D3-brane directions   
preserves supersymmetry since the Killing spinors do not depend on these 
isometric coordinates  \ci{KKU}.}
 to  the original 
D3-brane background and has the form 
\be
ds^2 = H^{1/2} ( d\tx^p d\tx^p + dy^t dy^t) \ , 
\ \ \ \ \
e^{\p} = H \ , \ \ \ \ \ \  C  = H^{-1}  \ , 
\ \ \ \ \   H= { \td R^4 \ov y^4} \ . 
\la{ins}
\ee
Note that this conformally flat $D=10$ string-frame  
metric
is actually  equivalent to the \ads metric \rf{met}:
changing the coordinates $y^t$ so that  the 
radial coordinate gets inverted, 
$y= 1/y'$,  we get
(we set $\td R=1$ as above)
\be
ds^2 = y'^2 d\tx^p d\tx^p +  { 1 \ov y'^2} dy'^t dy'^t \ , \ \ \ \ \ \
e^{\p}  = C^{-1} = y'^4 \ . 
\ee
Thus, like the action   \rf{SA},  the dual 
 action \rf{rsa} can    also  be 
{\it directly}  interpreted as  describing  a superstring propagating 
in \ads space, now 
 supplemented  not by the  4-form background  but by the 
 dilaton and 0-form backgrounds.

Since the fermionic term in \rf{rsa}
 does not depend on the 2-d metric, 
the  semiclassically equivalent Nambu-type action obtained 
by solving for  $g_{ij}$  is thus {\it also} 
quadratic in $\theta$:
$$
\td S = - { 1 \ov 2} \int d^2\sigma\ \bigg[ \sqrt {- \det \big[
{ 1\ov y^{2}}(\partial_i \tx^p \partial_j \tx^p +  \partial_i y^t
\partial_j y^t) \big] } $$
\be 
+\  4i \eps^{ij} \bar
\vt ( \del_i \tx ^p \Gamma^{p} + 
\partial_i y^t  \Gamma^t ) 
 \partial_j \vt \bigg]  \ . 
\la{rsan}
\ee
A possible  reparametrization gauge choice here 
may be the static gauge:
$\tx^i = \s^i, \ i=0,1$, leading to 
 the free  `kinetic'  $\bar \vt\del \vt$  term in the action ($\pi=2,3$)
$$
\td S = - { 1 \ov 2} \int d^2\sigma\  \bigg[ \  { 1 \ov y^2} \sqrt {- \det (\eta_{ij} + \partial_i \tx^\pi \partial_j \tx^\pi +  \partial_i y^t
\partial_j y^t)  } $$
\be 
+\  4i \eps^{ij} \bar
\vt   \Gamma_{i}   \partial_j \vt    
   +  
 4i \eps^{ij} \bar
\vt ( \del_i \tx ^\pi \Gamma^\pi + 
\partial_i y^t  \Gamma^t ) 
 \partial_j \vt   
   \bigg]  \ .
\la{rsany}
\ee
The semiclassical expansion is then developed by starting with a 
particular solution  for the string coordinates  (e.g., 
$y=y(\s), \ \tx^\pi=0$ as in \ci{MR}, see  below) 
and  integrating over small fluctuations near it.

Let us note that 
in  fixing the static gauge one assumes that the ground state 
of the string is massive, i.e. this gauge is 
appropriate for a solitonic
 string  or wound  string state  but 
cannot be used  to describe  a  spectrum 
of a fundamental string in  the zero-winding sector.
In flat space an adequate gauge for the latter  purpose is the 
combination of the conformal gauge $\sqrt{-g} g^{ij}= \eta^{ij}$ and 
the 
light-cone gauge $x^+ =\s^0$.
 However,  fixing the light-cone  gauge in a curved space 
which is not a direct product $R^{1,1} \times M^{D-2}$ 
may not always be possible (see, e.g., \ci{rudd}). 
Indeed, in  contrast to what happens in  flat space
(or in a plane-wave type 
backgrounds \ci{HS,RT}),  
in the present \ads  case  the conformal-gauge 
bosonic string equations of 
motion for $x^0, x^1$  corresponding to the metric \rf{met}\  
$ \del^i (y^{2} \del_i) x^p =0$
do not have $x^+ =\s^0$ as a solution for a generic 
classical  configuration 
of $y^t$  with $y=|y|$ satisfying
 $  \del^i \del_i \log y  
+ y^2 \del^i x^p  \del_i  x^p =0$.


4. \   In order to  achieve  a  better understanding 
of the  \ksym gauge choice in 
\rf{gau} it is useful to  study 
 the issue of invertibility
of the fermionic kinetic operator in the actions \rf{SA},\rf{rsa}.
In particular, we shall consider the 
flat space case obtained by omitting  
(or just treating  as constant)  the $y^2 $ and $1/y^2$ factors 
in the metrics \rf{met},\rf{ins} and the actions \rf{SA},\rf{rsa}.
We shall choose the  conformal gauge $\sqrt{-g}  g^{ij} = \eta^{ij}$.
In  general, 
the constraints  coming from the  equation of motion for the 2-d  metric
can be written  in terms of the vielbein components of 
the  `momentum'  $\Pi^{\hat a}_i$ 
defined by the $g_{ij}$-dependent 
part of the action  which does not include the 
 WZ term
 ($z,\bar z = \s \pm \tau, \ 
\s^0\equiv \tau, \ \s^1\equiv \s$)
\begin{equation}
 \Pi_z  \cdot \Pi_{ z}\equiv \Pi_z^p  \Pi_z^p   + \Pi_z^t \Pi_z^t   =   
0\ , \ \ \
\ \
\ \   \Pi_{\bar z} \cdot
\Pi_{ \bar z}\equiv\Pi_{\bar z} ^p  \Pi_{\bar z}^p  + \Pi_{\bar z}^t  
\Pi_{\bar
z}^t  =0 \  \la{con}
\end{equation}
Dots stand for the fermionic terms 
in the constraints and  as  above  the
 indices $p$ are contracted with 4-d
Minkowski metric  and  the indices $t$ --  with   6-d
Euclidean
metric. 
In the case of  the action  \rf{SA}\ 
$\Pi^p_i = y  ( \del_i x^p 
 - 2 i \bar \vt \Gamma^{p} \partial_j \vt) 
, \
\Pi^t_i =  y^{-1} \del_i y^t
.$ 

Before \ksym gauge fixing   
the quadratic fermionic terms in the flat-space GS action are 
\be
 \bar \theta^1 (\Pi \cdot \Gamma)_z
\partial_{\bar z }  \theta^1\equiv  \bar \theta^1 A_1  \theta^1
\ , \ \ \ \ \ \ \
 \bar \theta^2 (\Pi \cdot \Gamma)_{\bar z }
\partial_{ z }  \theta^2 \equiv  \bar \theta^2 A_2 \theta^2 \ . 
\ee
On  the classical equations and constraints we get 
$
 (A_1)^2 =  (A_2)^2 =0  ,$ 
i.e. 
the fermionic operator  is degenerate for any  classical 
string background. 
As we shall see below,  after the \ksym gauge fixing  in \rf{gau} 
the degeneracy is removed   provided the background
is constrained  in a certain way (e.g., to have  non-zero
$\Pi_\sigma^2$ for vanishing $\Pi_\tau ^2$ for massless 10-d states).

The  quadratic  term in  the fermionic  part of the
gauge-fixed action \rf{SA} is 
(we omit
the fermionic terms in $\Pi$)
\be
\bar \vt  \ y \left[(\Pi \cdot \Gamma)_z
\partial_{\bar z} + (\Pi \ast \Gamma)_{\bar z}
\partial_ z\right] \ \vt \ \equiv\  \  \bar \vt \   A \  \vt \ ,
\ee
where we introduced  the notation 
\be
\Pi^p_i  \Gamma^p + \Pi^t_i  \Gamma^t =
(\Pi \cdot
\Gamma)_i \ ,  \ \ \ \ \ \ \ \
\Pi^p_i  \Gamma^p   - \Pi^t_i  \Gamma^t =
(\Pi \ast
\Gamma)_i \ .
\ee
Using the equations of motion for $X^{\hat m}$ and 
 the constraints  \rf{con},  the square of the kinetic operator $A$ can be written as  
$$
A^2 = y^2  [ (\Pi \cdot \Gamma)_z (\Pi \ast \Gamma)_{\bar z} +
(\Pi \ast
\Gamma)_{\bar z}  (\Pi \cdot \Gamma)_z] \partial_{ z}   
\partial_{\bar z} + ...  
$$
\begin{equation}
= \ y^2 [ \Pi^p_z \Pi ^{p}_{ \bar z} - \Pi^t_z \Pi ^{t }_{\bar z} ]
\partial_{
z}  \partial_{\bar z} + ... \ , 
\end{equation}
where dots stand for lower-derivative 
 $\del y$ dependent terms 
which are absent in the   flat space limit  ($y=\const$).
In flat space  
 $A^2$ 
is invertible even on massless ($M_{10}^2 =0 $) 
 10-d string states with
$(\Pi \cdot
\Pi)_\tau=0$ and  $(\Pi\cdot \Pi)_\s=0$   if the  $X^{\hat m}$ 
background
is a BPS one, 
\begin{equation}
 (\Pi^p  \Pi^p)_\tau =- M_4^2\ , \ \ \   \qquad  (\Pi^t  \Pi^t)_\tau
= Z^2 \ ,\   \qquad M^2_4 = Z^2  \ , 
\end{equation}
\begin{equation}
A^2 = -2 y^2 M_4^2  \partial_{ z}  \partial_{\bar z} =  
-2y^2 Z^2
\partial_{
z}  \partial_{\bar z} \ . 
\end{equation}
In the case of the  dual action \rf{rsa} the constraints 
are expressed in terms of  
 $\Pi^p_i = y^{-1}  \del_i \tilde x^p 
 , $  and 
$\Pi^t_i = y^{-1} \del_i y      
$,  i.e. do not include  fermionic terms. 
In the flat space limit (i.e. ignoring derivatives of $y$)
the square of the corresponding 
fermionic kinetic operator takes the form 
\begin{equation}
 \td A^2   =  y^2 \Pi^2 \del_z \del_{\bar z} \ , \ \ \ \ \ \ \ 
\Pi^2 \equiv 
\eta_{\hat a \hat b} \Pi_z ^{\hat a} \Pi_{\bar z} ^{\hat b} \ . 
\ee
Thus for the
invertibility of $\td A^2$  we need 
$
\Pi ^2 = -  ( \Pi \cdot \Pi)_\tau +  (\Pi \cdot \Pi)_\sigma
= 2(\Pi \cdot \Pi)_\sigma
 \not=0 ,  $
 where we have used the constraint
$( \Pi \cdot \Pi)_\tau +  (\Pi \cdot \Pi)_\sigma =0$. 

Similar conclusion is reached in the curved space case.
In particular, 
it is  possible to show that the fermionic operator
$ \td A^2$ is degenerate  on a general  center-of-mass
($\s$-independent)  solution
of the  bosonic equations  and  constraints that follow from 
the dual  action \rf{rsa}.
 Indeed, if
we assume that $\del_\s \tx^p=0, \ \del_\s y^t=0$, then 
$\del_\tau \tx^p= v^p y^2 , \  \del_\tau y^t= u^t y^2$,  where 
$v^p,u^t$ are constant vectors subject to the constraint
\be
v^p v^p + u^t u^t =0   \ . \la{coo}
\ee
As a result,\foot{Similar solution
of string equations in $AdS_n \times S^n$  was found by 
R.R. Metsaev (unpublished).}   
\be
\tx^p= a^p + v^p f(\tau)   \ , \ \ \ \ \ \ \ \
y^t= b^t  + u^t f(\tau)\ ,
\ee
\be
  f(\tau) = - { b\cdot u \ov u^2}  + { \omega \ov u^2} \ 
 {\rm tan } \ \omega (\tau-\tau_0)\ , \  \ \ \ \ 
\omega^2 \equiv {  b^2 u^2 - (b\cdot u)^2 }  , 
\ee
where  the  constants $a^p,v^p,b^t,u^t,\tau_0$
are functions of the initial data         $\tx^p(0), \del_\tau \tx^p(0),$ $ 
y^t(0), \del_\tau y^t(0)$.  \foot{Note that 
$y^t$ depends only on part of $b^t $  which is transverse to
$u^t$, i.e.
$b^t  - { b\cdot u \ov u^2} u^t$. The parameter $\tau_0$  
compensates for the  `missing' longitudinal  component of $b^t$
and can be determined from the constraints
$u^t y^t(\tau_0) =0$ or 
     $y^t(0)  \del_\tau y^t(0) = - { \omega^3 \ov u^2 } { \sin \omega \tau_0 \ov  \cos^3 \omega \tau}$.}
The fermionic term in \rf{rsa} is  thus proportional  to 
$
y^2(\tau) \ \bar
 \vt ( v^p \Gamma^p + u^t  \Gamma^t ) \partial_\s \vt , $
and   is degenerate ($\td A^2=0$)   since 
$( v^p \Gamma^p + u^t  \Gamma^t )^2 =0$ as  follows from \rf{coo}.
To  have a non-degenerate
fermionic operator one  has  to consider 
inhomogeneous ($\s$-dependent)
string configurations.  

5. \ 
A particular class of such  static but
$\s$-dependent configurations  is, in fact, of physical interest in connection
with Wilson loop calculations in \ci{MR}. 
Let us now comment on the application of the \ads superstring 
action \rf{SA}
to this problem.
The aim is  to compute  the string path integral 
$
e^{-W} = \int [dxd\theta]  e^{-S}
$
by expanding near a 
 particular classical string solution.
The action $S$  may be  taken in the Nambu form (i.e.  with $g_{ij}$ 
 eliminated)\foot{The two forms of the string action 
are equivalent  in  this  context of semiclassical expansion \ci{FFT}.}
 and  one  may 
 fix the static gauge  $x^0 =  \s^0, \ \  x^1  =  \s^1$,
where $\s^i $ are 
the (Euclidean)
world-sheet coordinates  which run  from 0 to $T$ and $-L/2$ to $L/2$.
Following  \ci{MR},  let us    consider  a 
 string solution which is static 
and has only the radial coordinate  
 $y=\sqrt{y^ty^t}$ changing with $\s^1$. 
The  bosonic part of  string action  is  then proportional to 
$ T \int d\s^1  \sqrt {  (\del_1 y)^2 +  y^4}.$
The stationary point is determined by  the equation 
\be 
(\del_1 y)^2 +  y^4 = a y^8\ , \  \  \ \ \ \ \ a=\const \ ,  \la{eqq}
\ee
 with the boundary condition  that  $y$ takes  the 
minimal value $y_*=a^{-1/4}$ 
 at $\s=0$ and runs to  infinity at $-L/2$  and  $L/2$.  While 
 the solution $y(\s^1)$  does not have a 
 simple analytic form, \rf{eqq}   allows to  
eliminate $\del_1 y $ in terms of $y$. 
The classical value of the string action is  proportional to 
$ T/L$.  Expanding  the action near  the solution
one may  compute the 1-loop correction to the effective potential.

 Here we shall consider only  the quadratic
fermionic part as it follows from  the Nambu  analogue of the gauge-fixed action
\rf{SA}.
Using  \rf{eqq}, one finds 
that the classical  value of the induced Euclidean 
 2-d metric is 
$g_{ij}= y^2  \delta_{ij} + { 1 \ov y^{2}} \del_i y \del_j y=
\diag( y^2,  a y^6)$. Thus 
 $ \sqrt g g^{ij} =\diag (\sqrt a y^2, { 1 \ov  \sqrt a y^2}) ,  $  
and 
  the sum of the quadratic  fermionic terms in the action   takes the form
$$
S_{F}
 = 2i \int d^2\sigma\ \bigg(
\sqrt g g^{ij} y^2  
     \bar \vt \Gamma_i \partial_j\vt 
   -   \eps^{ij}   \partial_i y^t \bar \vt \Gamma^t
\partial_j\vt \ \bigg) $$
\be
 = 2i \int d^2\sigma\ \bar  \vt  \bigg [  { 1 \ov \sqrt a} 
\Gamma^1 \partial_1  +  
(
\sqrt a  y^4   \Gamma^0  +   \partial_1 y \G^y)  \partial_0   
  \bigg]  \vt   \ . 
\label{SF}
\ee
We used the fact  that 
 in the static gauge 
 $\del_i x^p = \delta^p_i$
 and took 
 into account that only the radial part of $y^t$ has  a 
non-trivial $\s^1$-dependent  background value
(the radial component  $\G^y$ depends on constant angular
 parameters).  
The  resulting fermionic operator is non-degenerate,
 and it may be possible
to  compute  the  contribution  of its determinant to 
$W$  in  the leading order in  large $L,T$.

In flat space  (or  for a  $y$=const solution) 
one finds that 
the fluctuations of $d-2=8$  transverse  bosonic coordinates
 (which are  periodic in $\s^0$  and $\s^1$)
give \ci{LU}\ \ \ 
$ W= \ha (d-2) \log \det (- \partial^2)  
 =  - (d-2) {\pi T\ov 24 L}.$
In  the flat-space   superstring case this  contribution 
 is cancelled by the contribution of 
the fermionic determinant:   the total effective number of 
transverse 
world-sheet degrees of freedom is equal to zero because of supersymmetry.
It  is plausible  that 
the  1-loop  $1/L$ correction to the effective potential 
will not, however, vanish in
 the present curved space case 
 as  there is no reason to 
expect that the  solution  $y=y(\s^1)\not=\const$  should  preserve  
 effective world-sheet  supersymmetry.

6. \ To conclude, we have shown that choosing  the 
`D3-brane' or `4-d space-time' adapted  \ksym gauge in the \ads 
superstring action  and   duality-rotating the four   isometric
 space-time coordinates  one obtains  an action  in which 
the fermionic term is  quadratic  and does not depend
on  the world-sheet metric.  The  `4+6'  Cartesian 
parametrisation of the 10-d space  thus led to a  substantial 
simplification   of the fermionic sector of the theory. 
This 
should hopefully allow one 
to make progress towards  extracting more non-trivial 
information  about  the \ads string theory and thus 
about  its dual \ci{MA} -- $N=4$  super Yang-Mills theory.


\bigskip

We are  grateful to   J. Rahmfeld,  L. Susskind  and, especially, 
R.R. Metsaev 
for  useful explanations and  discussions.
 The  work of R.K  is supported by the NSF grant PHY-9219345.
The   work of A.T.  is supported in part
by PPARC,   the European
Commission TMR programme grant ERBFMRX-CT96-0045
and  the INTAS grant No.96-538.


\end{document}